\documentclass[preprint,tightenlines,eqsecnum,floats,aps,amsmath,amssymb,nofootinbib,prd,showpacs]{revtex4}

\usepackage{amssymb}
\usepackage{stmaryrd}
\usepackage{amsmath}
\usepackage{amsfonts}
\usepackage{mathrsfs}
\usepackage{CJK}
\usepackage{amsmath,amssymb,amsfonts}
\usepackage{graphicx}
\usepackage{subfigure}
\def\be{\begin{equation}}
	\def\ee{\end{equation}}
\def\ba{\begin{eqnarray}}
	\def\ea{\end{eqnarray}}
\def\nn{\nonumber}

\begin{document}
%\date\today
%\preprint{????}

\title{Collisional Penrose process with spinning particles in braneworld black hole}

\author{Yongbin Du}
\affiliation{Department of Physics, South China University of Technology, Guangzhou 510641, China}

\author{Yunlong Liu}
\affiliation{Department of Physics, South China University of Technology, Guangzhou 510641, China}

\author{ Xiangdong Zhang\footnote{Corresponding author. scxdzhang@scut.edu.cn}}
\affiliation{Department of Physics, South China University of Technology, Guangzhou 510641, China}

\date{\today}

%\date{\today}

\begin{abstract}
The Penrose process of an extremal braneworld black hole is studied.
We analyze the Penrose process by two massive spinning particles collide near the horizon. By calculating the maximum energy extraction efficiency of this process, it turns out that the maximal efficiency increases as the tilde charge parameter $d$ of the braneworld blackhole decreases. Interestingly, for the negative value of $d$, the efficiency can be even larger than the Kerr case.

\end{abstract}
\maketitle
%%%%%%%%%%%%%%%%%%%%%%%%%%%%%%%%%%%%%%%%%%%%%%%%%%%%%%%%%%%%%%%%%%%%%
\section{INTRODUCTION}
	In the ergosphere of Kerr black hole, there exist some time-like orbits whose energy can be strangely negative due to the Killing vectors in this region being always spacelike. Penrose first conceived the concept of extracting energy from these special orbits\cite{penrose}. His creative scenario alleged that if an object plunging into the ergosphere of a rotating black hole and breaks into two pieces and one of them falls into the event horizon with negative energy, the other one can escape from the black hole, carrying more energy than the original object. This mechanism is now well known as the Penrose process. Although some early work showed \cite{Bardeen:1972fi, Wald:1974kya} the process may rarely happen in astrophysics. However, soon after that, the scenario of collisional Penrose process was proposed \cite{collision} and was highly expected at one time to acquire energy from Kerr black holes with greater efficiency. Nevertheless, some researches such as\cite{Piran:1977dm} uncovered that it may not get higher efficiency as expected, making this interesting topic gradually recede from the physicists' arena. It did not get its rebirth until Banados, Silk and West $($BSW$)$ \cite{Patil:2011yb} find that the center-of-mass energy can be arbitrarily high when the collision happens near the horizon and the incident particles have tuned angular momentum. Therefore, the collisional Penrose process based on the BSW mechanism has been constructed and found to have much higher efficiency than the previous version \cite{Harada:2012ap,Bejger:2012yb,Schnittman:2014zsa}. These results were soon generalized to the non-equatorial plane\cite{Leiderschneider:2015kwa}. Moreover, the BSW effect including spinning particles are discussed in \cite{Zaslavskii:2010jd, Deriglazov:2018vwa, Zaslavskii:2016dfh, Wei:2010vca, Harada:2011xz, Kimura:2010qy,Guo:2016vbt}. Then, the collisional Penrose process has also been generalized to the spinning particles case\cite{Maeda:2018hfi, Okabayashi:2019wjs}, It turns out that when the spinning particles were considered, the efficiency of contracting energy from the blackhole was greatly improved compared with the spinless case \cite{Maeda:2018hfi}. In particular, Maeda, Okabayashi and Okawa calculate spinning test particles colliding near the horizon of the Kerr background and obtain their maximal efficiency, which is defined by $($input energy$)/($output energy$)$, is around 15.01 for massive particles\cite{Maeda:2018hfi}.

	On the other hand, most of the previous studies on the Penrose process were focus on four dimensions. However, whether our spacetime has extra dimensions is a fundamental question in modern physics. In past one hundred years different types of extra dimensional models were proposed, such as Kaluza-Klein $($KK$)$ model \cite{kaluza,klein} and string/M theory which assumes that the extra dimension is needed but quite tiny(approach Planck scale) in order to explain why it is not found in experiments. While braneworld model\cite{model1} conceives that our visible universe is localized on a 3-brane in high dimension space-time and the extra dimensions are large. Randall and Sundrum developed double brane models based on Arkani-Hamed, Dimopoulos and Dvali's $($ADD$)$ theory \cite{Arkani-Hamed, Dimopoulos, Dvali} and tactfully solved the hierarchy problem\cite{model1}. Later, they further developed their model in which all the matter of our universe is localized on the single brane of original point$(\phi=0)$ and the extra dimension is infinite\cite{model2}. Interestingly, the blackhole solution in braneworld model is obtained by Naresh Dadhich $et$ $al.$ gives an effective solution of rotating black hole, which is the form of Reissner-Nordstrom metric but the real charge is replaced by a new parameter tidal charge $d$ as an imprint of extra dimension\cite{Dadhich:2000am}. After their vital work,  Aliev and Gumrukcuoglu acquired stationary and axisymmetric solutions describing charged rotating black holes on a 3-brane in the Randall-Sundrum braneworld\cite{Aliev:2005bi}. The Kerr solution then can be viewed as the limit of the tidal charge $d=0$. The collisional Penrose process of braneworld Kerr black hole with spinless particles has been discussed in\cite{Khan:2019gco}. The present paper aims to study the collisional Penrose process with spinning particles. The necessities for such extension are two folds, first, the spinning massive particles are more realistic than the massive spinless particles. Second, the spinning particles in the Penrose process usually have much higher maximal energy contraction efficiency compared with the spinless particles \cite{Okabayashi:2019wjs}.

	This paper is organized as follows. After an introduction, in section 2, we give the equations of motion of spinning test particles in braneworld Kerr black hole. In section 3, constraints on the orbit are obtained. Based on these constraints we discuss the collisional process near the horizon in section 4. The maximal efficiency of extracting energy in braneworld black hole with different tidal charge $d$ is given in section 5 and the concluding remarks are presented in section 6.

	%%%%%%%%%%%%%%%%%%%%%%%%%%%%%%%%%%%%%%%%%%%%%%%%%%%%%
	%%%%%%%%%%%%%%%%%%%%%%%%%%%%%%%%%%%%%%%%%%%%%%%%%%%%%
	\section{EQUATIONS OF MOTION}
	In this section, we will give the differential equations of motion for spinning test particles moving in braneworld Kerr spacetime.
	%%%%%%%%%%%%%%%%%%%%%%%%
	\subsection{Braneworld Kerr space-time}
	The metric of braneworld Kerr spacetime reads \cite{Dadhich:2000am}:
	\begin{equation}
		g_{\mu \nu}=\left(
		\begin{array}{cccc}
			-\frac{\Delta -a^2 \sin ^2\theta}{\Sigma } & 0 & 0 & -\frac{a \sin ^2\theta \left(\left(a^2+r^2\right)-\Delta \right)}{\Sigma } \\
			0 & \frac{\Sigma }{\Delta } & 0 & 0 \\
			0 & 0 & \Sigma  & 0 \\
			-\frac{a \sin ^2\theta \left(\left(a^2+r^2\right)-\Delta \right)}{\Sigma } & 0 & 0 & \frac{\sin ^2\theta \left(\left(a^2+r^2\right)^2-a^2 \Delta  \sin ^2\theta \right)}{\Sigma } \\
		\end{array}
		\right),
	\end{equation}
	where
	\begin{eqnarray}
		\Delta &=&a^2+d-2 M r+r^2,\\
		\Sigma &=&a^2 \cos ^2\theta+r^2,
	\end{eqnarray} where $d$ is the tidal charge parameter which could be either positive or negative. When this parameter $d=0$, the above metric reduces to the Kerr metric. We should point out that this metric can be seen as the Kerr-Newman solution if we replace the tidal charge parameter $d$ with $Q^2$. However, $d$ is not always positive which, as we will show in this paper, plays an important role in achieving higher efficiency of the collisional Penrose process. To make sure the horizon exists, the following condition must be satisfied
	\begin{eqnarray}
		M^2 \geq a^2+d.
	\end{eqnarray}
	The equal is taken only in the case that it is extreme braneworld black hole. Besides, it is evidently that when $d$ is negative in the extreme braneworld black hole, $a$ can be larger than $M$ to make sure the event horizon exists. The radius of the outer horizon reads
	\begin{eqnarray}
		r_{H}=\sqrt{-a^2-d+M^2}+M.
	\end{eqnarray}

	%%%%%%%%%%%%%%%%%%%%%%%
	\subsection{Equation of spinning test particles}
	Due to the existence of spin, the orbit of spinning particles is described by MPD equations rather than geodesics \cite{Papapetrou1951Spinning,Dixon1970Dynamics1,Dixon1970Dynamics2}

	\begin{eqnarray}
		\frac{Dp^{\mu}}{d\tau}&=&	-\frac{1}{2}R^{\mu}_{\nu \rho \sigma}v^{\nu}S^{\rho \sigma},\\
		\frac{DS^{\mu \tau}}{d\tau}&=&p^{\mu}v^{\nu}-p^{\nu}v^{\mu},
	\end{eqnarray}
	where $p^{\mu}$, $v^{\mu}=dz^{\mu}/d\tau$ and $S^{\mu \nu}$ are the 4-momentum, 4-velocity and the spin tensor respectively. A set of supplementary conditions are also needed \cite{Wald1972Gravitational}
	\begin{eqnarray}
		S^{\mu \nu}p_{\nu}=0.
	\end{eqnarray}
	Moreover, the mass of particles is defined by
	\begin{equation}
		\mu^2=-p^{\mu}p_{\mu}.
	\end{equation}
	We use a specific 4-momentum as well, which reads
	\begin{equation}
		u^{\mu}=\frac{p^{\mu}}{\mu}.
	\end{equation}
	We should also normalize the affine parameter as
	\begin{equation}
		u^{\mu}v_{\mu}=-1.
	\end{equation}
	Hence we could get the difference between 4-momentum and 4-velocity as follows \cite{Motoyuki_1998}
	\begin{eqnarray}
		v^{\mu}-u^{\mu}=\frac{S^{\mu \nu}R_{\nu \rho \lambda}u^{\rho}S^{\sigma \lambda}}{2\mu^2+\frac{1}{2}R_{\alpha \beta \gamma \delta}S^{\alpha \beta}S^{\gamma \delta}}.
	\end{eqnarray}

	%%%%%%%%%%%%%%%%%%%%%%%%%
	\subsection{Conservation quantities}
	As we have said in the preceding text, the world line of spinning particles is not geodesics anymore, causing the necessary modification of the definition of conserved quantities \cite{Hojman_1978}
	\begin{equation}
		Q_{\xi}=p^{\mu}\xi_{\mu}+\frac{1}{2}S^{\mu \nu}\nabla_{\mu}\xi_{\nu}.
	\end{equation}
	In braneworld Kerr space-time, we invariably have two Killing vectors. By introducing a tetrad basis as
	\begin{eqnarray}
	\label{Tetrad1}
	&&e^{(0)}_{a}=\frac{\sqrt{\Delta }}{\sqrt{\Sigma }}dt_{a}-\frac{\sqrt{\Delta } \left(a \sin ^2(\theta )\right)}{\sqrt{\Sigma }}d\varphi_{a},\notag\\
	\label{Tetrad2}
	&&e^{(1)}_{a}=\frac{\sqrt{\Sigma }}{\sqrt{\Delta }}dr_{a},\\
	\label{Tetrad3}
	&&e^{(2)}_{a}=\sqrt{\Sigma }d\theta_{a},\\
	\label{Tetrad4}
	&&e^{(3)}_{a}=-\frac{a \sin (\theta )}{\sqrt{\Sigma }}dt_{a}+\frac{\left(a^2+r^2\right) \sin (\theta )}{\sqrt{\Sigma }}d\varphi_{a},
	\end{eqnarray}
	the Killing vectors can be written as
	\begin{eqnarray}
		\xi^{(t)}_{\mu}&=&-\sqrt{\frac{\Delta}{\Sigma}}e^{(0)}_{\mu}-\frac{a\sin{\theta}}{\sqrt{\Sigma}}e^{(3)}_{\mu},\\
		\xi^{(\varphi)}_{\mu}&=&a\sqrt{\frac{\Delta}{\Sigma}}\sin^2{\theta}e^{(0)}_{\mu}+\frac{(r^2+a^2)\sin{\theta}}{\sqrt{\Sigma}}e^{(3)}_{\mu}.
	\end{eqnarray}
	From it, we can easily acquire the energy $E$ and the $z$ component of the angular momentum $J$ in the equatorial plane $(\theta=\pi/2)$
	\begin{eqnarray}
		E&=&p^{(0)} \frac{\sqrt{\Delta}}{r}+ p^{(3)}\frac{a}{r}+S^{(0)(1)}\frac{(d-M r)}{r^3},\\
		J&=&p^{(0)}\frac{a  \sqrt{\Delta}}{r}+ p^{(3)}\frac{ \left(a^2 r^2+r^4\right)}{r^3}\notag\\
		&&+S^{(1)(3)}\frac{ \sqrt{\Delta}}{r}+ S^{(0)(1)}\frac{a (d-r (M+r))}{r^3}.
	\end{eqnarray}
	%%%%%%%%%%%%%%%%%%%%%%%%%%%%%
	\subsection{Equations of motion in the equatorial plane}
	In this paper, we merely discuss the orbit in the equatorial plane, which means $\theta=\pi/2$ and we introduce a spin vector \cite{Maeda:2018hfi}
	\begin{equation}
		s^{(a)}=-\frac{1}{2\mu}\epsilon^{(a)}_{(b)(c)(d)}u^{(b)}S^{(c)(d)},
	\end{equation}
	where $\epsilon$ is the antisymmetric tensor with $\epsilon_{(0)(1)(2)(3)}=1$. We also find all the components of the spin vector is trivial except
	\begin{equation}
		s^{(2)}=-s.
	\end{equation}
	Therefore it can be obtained that
	\begin{equation}
		S^{(\mu) (\nu)}=\left(
		\begin{array}{cccc}
			0 & -p^{(3)} s & 0 & p^{(1)} s \\
			p^{(3)} s & 0 & 0 & p^{(0)} s \\
			0 & 0 & 0 & 0 \\
			-p^{(1)} s & -p^{(0)} s & 0 & 0 \\
		\end{array}
		\right).
	\end{equation}
	Thus the conserved quantities can be written as
	\begin{eqnarray}
		E&=&\frac{p^{(0)} r^2 \sqrt{\Delta}+ p^{(3)} (a r^2-d s+M r s)}{r^3},\\
		J&=&\frac{p^{(0)}( r^2 s \sqrt{\Delta}+a r^2 \sqrt{\Delta})}{r^3}\notag \\
		&&+\frac{ p^{(3)} (a^2 r^2+a s (r (M+r)-d)+ r^4)}{r^3}.
	\end{eqnarray}
	By solving these equations we obtain
	\begin{eqnarray}
		u^{(0)}&=&\frac{E \left(-a d r s+a M r^2 s+a r^3 (a+s)+r^5\right)-J \left(r^2 (a r+M s)-d r s\right)}{\mu  \sqrt{a^2+d+r (r-2 M)} \left(d s^2-M r s^2+r^4\right)},\\
		u^{(3)}&=&\frac{r^3 (J-E (a+s))}{\mu  \left(d s^2-M r s^2+r^4\right)}.
	\end{eqnarray}
	For the reason of normalization condition
	\begin{equation}
		-(u^{(0)})^{2}+(u^{(1)})^{2}+(u^{(3)})^{2}=-1,
	\end{equation}
	we have
	\begin{equation}
		u^{(1)}=\sigma\sqrt{(u^{(0)})^{2}-(u^{(3)})^{2}-1},
	\end{equation}
	where $\sigma=\pm1$. The value is positive when the particle is outgoing and negtive when it is ingoing respectively. It is obtained hereinabove that
	\begin{eqnarray}
		v^{(0)}&=&\frac{\lambda_{2} u^{(0)}}{\lambda_{1}},\\
		v^{(1)}&=&\frac{\lambda_{2} u^{(1)}}{\lambda_{1}},\\
		v^{(3)}&=&\frac{u^{(3)} (\lambda_{2}+\lambda_{3})}{\lambda_{1}},
	\end{eqnarray}
	where
	\begin{eqnarray}
		\lambda_{1}&=&d s^2 \left(4 (u^{(3)})^2+1\right)-M r s^2 \left(3 (u^{(3)})^2+1\right)+r^4,\\
		\lambda_{2}&=&d s^2-M r s^2+r^4,\\
		\lambda_{3}&=&3 M r s^2-4 d s^2.
	\end{eqnarray}
	By using the transformation of coordinates Eqs.\eqref{Tetrad1}-\eqref{Tetrad4}, we finally get the differential equations
	\begin{eqnarray}
			\frac{\text{d}t}{\text{d$\tau $}}&=&\frac{u^{(0)} \left(a^2 \sqrt{\Delta} \left(d s^2-M r s^2+r^4\right)+r^2 \sqrt{\Delta} \left(d s^2-M r s^2+r^4\right)\right)}{r \Delta \left(d s^2 \left(4 (u^{(3)})^2+1\right)-M r s^2 \left(3 (u^{(3)})^2+1\right)+r^4\right)}\notag\\
			&&+\frac{u^{(3)} \left(a^3 \left(-3 d s^2+2 M r s^2+r^4\right)-a r (2 M-r) \left(-3 d s^2+2 M r s^2+r^4\right)\right)}{r \Delta \left(d s^2 \left(4 (u^{(3)})^2+1\right)-M r s^2 \left(3 (u^{(3)})^2+1\right)+r^4\right)},\\
		\frac{\text{d}r}{\text{d$\tau $}}&=&\frac{u^{(1)} \sqrt{\Delta} \left(d s^2-M r s^2+r^4\right)}{r \left(d s^2 \left(4 (u^{(3)})^2+1\right)-M r s^2 \left(3 (u^{(3)})^2+1\right)+r^4\right)},\\
			\frac{\text{d$\varphi $}}{\text{d$\tau $}}&=&\frac{a u^{(0)} \sqrt{\Delta} \left(d s^2-M r s^2+r^4\right)}{r \Delta \left(d s^2 \left(4 (u^{(3)})^2+1\right)-M r s^2 \left(3 (u^{(3)})^2+1\right)+r^4\right)}\notag\\
			&&+\frac{u^{(3)} \left(a^2 \left(-3 d s^2+2 M r s^2+r^4\right)-r (2 M-r) \left(-3 d s^2+2 M r s^2+r^4\right)\right)}{r \Delta \left(d s^2 \left(4 (u^{(3)})^2+1\right)-M r s^2 \left(3 (u^{(3)})^2+1\right)+r^4\right)}.
	\end{eqnarray}
	In what follows, for convenience of calculation, we rescale the variables to the dimensionless variables as
\begin{eqnarray}
\bar E=\frac{E}{\mu},\bar J=\frac{J}{\mu M},\bar s=\frac{s}{\mu M},\bar t=\frac{t}{\mu M},\notag \quad\bar r=\frac{r}{M},\quad\bar a=\frac{a}{\mu M},\quad \bar \tau=\frac{\tau}{\mu M}.
\end{eqnarray}
For simplicity, we omit the sign $"-"$ in the following discussion. Now the differential equations can be rewritten as
	\begin{eqnarray}
			\frac{\text{d}t}{\text{d$\tau $}}&=&\frac{u^{(0)} \left(a^2 \sqrt{\Delta} \left(d s^2-r s^2+r^4\right)+r^2 \sqrt{\Delta} \left(d s^2-r s^2+r^4\right)\right)}{r \Delta \left(d s^2 \left(4 (u^{(3)})^2+1\right)-r s^2 \left(3 (u^{(3)})^2+1\right)+r^4\right)}\notag\\
			&&+\frac{u^{(3)} \left(a^3 \left(-3 d s^2+2 r s^2+r^4\right)-a r (2 -r) \left(-3 d s^2+2 r s^2+r^4\right)\right)}{r \Delta \left(d s^2 \left(4 (u^{(3)})^2+1\right)- r s^2 \left(3 (u^{(3)})^2+1\right)+r^4\right)},\\
		\frac{\text{d}r}{\text{d$\tau $}}&=&\frac{u^{(1)} \sqrt{\Delta} \left(d s^2- r s^2+r^4\right)}{r \left(d s^2 \left(4 (u^{(3)})^2+1\right)- r s^2 \left(3 (u^{(3)})^2+1\right)+r^4\right)},\\
			\frac{\text{d$\varphi $}}{\text{d$\tau $}}&=&\frac{a u^{(0)} \sqrt{\Delta} \left(d s^2- r s^2+r^4\right)}{r \Delta \left(d s^2 \left(4 (u^{(3)})^2+1\right)- r s^2 \left(3 (u^{(3)})^2+1\right)+r^4\right)}\notag\\
			&&+\frac{u^{(3)} \left(a^2 \left(-3 d s^2+2 r s^2+r^4\right)-r (2 -r) \left(-3 d s^2+2 r s^2+r^4\right)\right)}{r \Delta \left(d s^2 \left(4 (u^{(3)})^2+1\right)- r s^2 \left(3 (u^{(3)})^2+1\right)+r^4\right)}.
	\end{eqnarray}
	%%%%%%%%%%%%%%%%%%%%%%%%%%%%%%%%%%%%%%%%%%%%%%%%%%%%%
	%%%%%%%%%%%%%%%%%%%%%%%%%%%%%%%%%%%%%%%%%%%%%%%%%%%%%
	
\section{CONSTRAINS OF THE ORBIT}
	%%%%%%%%%%%%%%%%%%%%%%
	\subsection{Different types of orbits}
\begin{figure}
	\centering
	\includegraphics{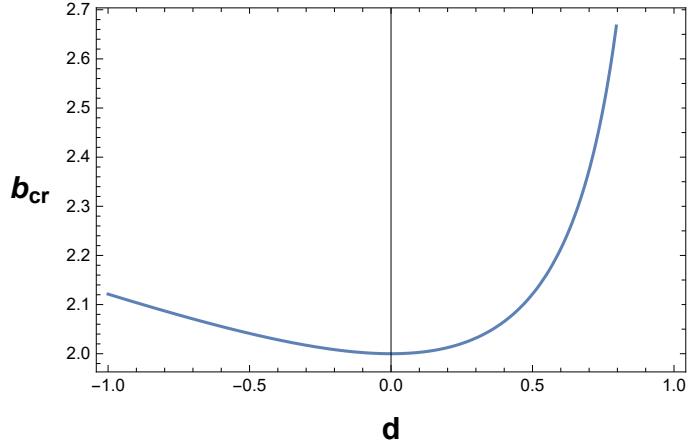}
	\caption{ The trend of $b_{cr}$ with parameter $d$ in extreme braneworld Kerr black hole. }
	\label{qq}
	\end{figure}

	Given we are exploring the collisional Penrose process where certain particles should escape with more energy after the collision in the ergosphere, it is essential to find the radial turning point, which means $\text{d}r/\text{d}t=0$. The differential equation with dimensionless variables makes it clear that the radial velocity of the particle depends on $E,J,s,r,a$ and $d$. Thus by taking distinct value of the turning point $r_{0}$, we could classify different types of orbits and obtain the relations between these quantities. By introducing the impact parameter $b=J/E$. When $\text{d}r/\text{d}t=0$, the radial function can be reduced to
	\begin{eqnarray}
			&&\left[E \left(-\frac{a d s}{r}+a r (a+s)+a s+r^3\right)-b E \left(a r-\frac{d s}{r}+s\right)\right]^2\notag\\
			&&-\Delta  r^2 (b E-E (a+s))^2+\Delta  \left(-P^2\right) r^4=0,
	\end{eqnarray}
	where
	\begin{equation}
		P=\frac{d s^2}{r^4}-\frac{s^2}{r^3}+1.
	\end{equation}
	We focus on the orbits whose turning points are exactly on the horizon $r_{H}$ which corresponding to the highest energy contraction efficiency \cite{Maeda:2018hfi,r6}. These orbits are usually called critical orbits \cite{Maeda:2018hfi,r6}. Then the critical value of impact parameter $b=\frac{J}{E}$ reads
	\begin{eqnarray}
			b_{cr}&=&\frac{a s \left(\left(\sqrt{-a^2-d+1}+1\right)^2+\sqrt{-a^2-d+1}-d+1\right)}{s \left(\sqrt{-a^2-d+1}-d+1\right)+a \left(\sqrt{-a^2-d+1}+1\right)^2}\notag\\
			&&+\frac{\left(\sqrt{-a^2-d+1}+1\right)^4+a^2 \left(\sqrt{-a^2-d+1}+1\right)^2}{s \left(\sqrt{-a^2-d+1}-d+1\right)+a \left(\sqrt{-a^2-d+1}+1\right)^2}.
	\end{eqnarray}
	In order to reach the horizon, the particles must have $b<b_{cr}$. Moreover, we can also see that if $d=0$, $b_{cr}$ comes back to the form in Kerr black hole \cite{Maeda:2018hfi}. In extreme braneworld black hole case, the above relation can be further reduced to
	\begin{equation}
		b_{cr}=\frac{a^2+1}{a}=\frac{2-d}{\sqrt{1-d}},
	\end{equation}
	which does not depend on $s$. The plot of $b_{cr}$ depends on parameter $d$ in extreme braneworld black hole is showed in Fig. \ref{qq}. We found that when $d$ is negative, we can see that the rate of change is not very large. While when $d=0$, $b_{cr}$ approaches the minimum value$(b_{cr}=2)$.%%%%%%%%%%%%%%

%We take $E=1$ to show the relation of $J,a,s$ and $r_{0}$ in Fig 2.%%%%%%%%%%%
	%\begin{figure}[t]
	%\centering
	%\includegraphics{J1.eps}
	%\end{figure}

	If the orbit is near critical$(b\approx b_{cr})$ but $b>b_{cr}$, the particle will bounce off near the horizon. In this paper, we consider the positive $J$ for it gives higher efficiency \cite{Maeda:2018hfi}. Moreover, we can also find that for non-critical orbit, the particle has no turning point and would plunge into the horizon \cite{Maeda:2018hfi,r6}.
	%%%%%%%%%%%%%%%%%%%%%%%%%%
	\subsection{Time-like condition of the orbits}
	To make sure the orbit exists near the horizon, the time-like condition should be imposed
	\begin{equation}
		-(v_{0})^2+(v_{1})^2+(v_{3})^2<0.
	\end{equation}
	Follow the same line of \cite{Maeda:2018hfi}, the above timelike condition can be reduced to
	\begin{equation}
		\label{timelikeEq}
		s^2 (a E+E s-J)^2<G,
	\end{equation}
where
	\begin{equation}
	\label{Gfunction}
	G=\frac{\left(d s^2+r^4-r s^2\right)^4}{r^6 \left(8 d^2 s^2-2 d \left(4 r^4+5 r s^2\right)+3 r^2 \left(2 r^3+s^2\right)\right)}.
	\end{equation}

\begin{figure}
	\centering
	\includegraphics{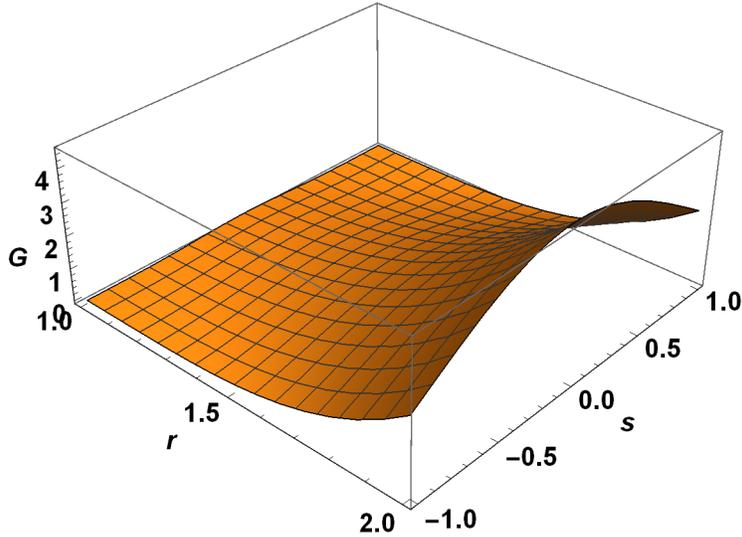}
\caption{ The image of the function $G$, where $d=-0.28$ is chosen.
}
	\label{timelike}
	\end{figure}

	The right of the inequality is a function of $r,d$ and $s$, and we show the image of the function $G$ in  Fig. \ref{timelike}.

\begin{figure}
		\centering
		\includegraphics{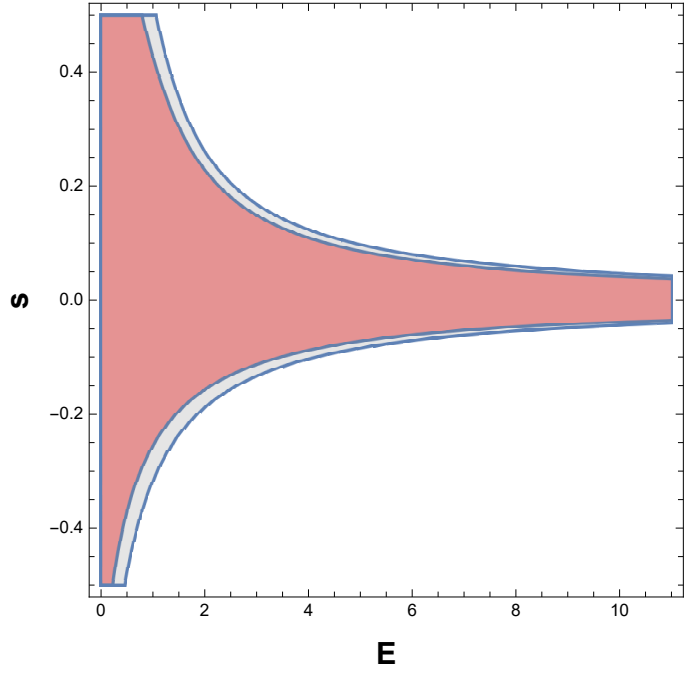}
		\caption{Allowed region of spin $s$ when $d=-0.24$$($ red $)$  and $d=0.08$$($ gray $)$.}
		\label{newconstrain}
	\end{figure}

	In the following, we focus on the extreme braneworld black hole corresponding to the highest energy extraction efficiency. In this case, we have $d=1-a^2$. Firstly, we shall take a view of the time-like constraints on the critical orbits$(b=b_{cr})$ under this background. We have
	\begin{equation}
		E^{2}<\frac{\left(\left(a^2-1\right) s^2-r^4+r s^2\right)^4}{r^6 s^2 \left(4 a^2+3 r-4\right) (a-b+s)^2 \left(2 \left(a^2-1\right) s^2+2 r^4+r s^2\right)}.
	\end{equation}
	If the particle comes from infinity, $E\geqslant1$, we could get allowed region of spin $s$ which shown in Fig. \ref{newconstrain}.

	If the orbit is non-critical, the time-like condition Eq.\eqref{timelikeEq} gives constraints on $b$
	\begin{equation}
		a-\frac{Y}{E}+s<b<a+\frac{Y}{E}+s,
	\end{equation}
	where
	\begin{equation}
		Y=\frac{\left(d s^2+r^4-r s^2\right)^2}{r^3 \sqrt{s^2 \left(8 d^2 s^2-8 d r^4-10 d r s^2+6 r^5+3 r^2 s^2\right)}}.
	\end{equation}
	We set $b=b_{cr}(1+\zeta)$ with $\zeta<0$, then we have
	\begin{equation}
		\frac{a-b_{cr}-\frac{Y}{E}+s}{b_{cr}}<\zeta <\frac{a-b_{cr}+\frac{Y}{E}+s}{b_{cr}},\label{xi}
	\end{equation}
	which gives the constraints on the impact parameter.
	%%%%%%%%%%%%%%%%%%%%%%%%%%%%%%%%%%%%%%%%%%%%%%%%%%%%%%%%
	%%%%%%%%%%%%%%%%%%%%%%%%%%%%%%%%%%%%%%%%%%%%%%%%%%%%%%%%
	\section{COLLISION OF MASSIVE SPINNING PAITICLES}
	Now we discuss the collision of massive spinning particles in the extreme braneworld black hole in order to get higher efficiency. The mass of the two colliding particles are chosen to be identical for simplicity and they are both from infinity. We use 1 and 2 to label them respectively. After their plastic collision, particle
3, whose 4-momenta is $p_3^\mu$, will move to infinity, while the particle 4 with $p_4^\mu$ is going to fall into the black hole \cite{Maeda:2018hfi,r6}. By assuming
	\begin{gather}
		p^{\mu}_{1}+p^{\mu}_{2}=p^{\mu}_{3}+p^{\mu}_{4},\\
		S^{\mu \nu}_{1}+S^{\mu \nu}_{2}=S^{\mu \nu}_{3}+S^{\mu \nu}_{4}.
	\end{gather}
	which means the total momentum and spins are invariable, and the total energy and angular momentum are conserved during the collision,
	\begin{gather}
		E_{1}+E_{2}=E_{3}+E_{4},\\
		J_{1}+J_{2}=J_{3}+J_{4}.
	\end{gather}
	To make it easier to discuss the quantities of the orbits at the plastic collision, we simplify the conservative relations as follow
	\begin{gather}
		E_{1}+E_{2}=E_{3}+E_{4},\\
		J_{1}+J_{2}=J_{3}+J_{4},\\
		s_{1}+s_{2}=s_{3}+s_{4},\\
		u^{(1)}_{1}+u^{(1)}_{2}=u^{(1)}_{3}+u^{(1)}_{4}.\label{radialv}
	\end{gather}
	We also assume the collision point is quite close to the horizon$(r_{c}=\frac{1}{(1-\epsilon)},0<\epsilon<<1)$, as is expected to acquire higher energy extraction efficiency. Thus the radial velocity can be expanded as
	\begin{equation}
		u^{(1)}=\sigma\frac{E+a^2 E-a J}{\left| a s-1\right|\epsilon}+O(\epsilon^0).
	\end{equation}
	Given the conservation equation of radial velocity \eqref{radialv}, we have
	\begin{eqnarray}
			&&\sigma_{1}\frac{E_{1}+a^2 E_{1}-a J_{1}}{\left|a s_{1}-1\right|\epsilon}+\sigma_{2}\frac{E_{2}+a^2 E_{2}-a J_{2}}{\left| a s_{2}-1\right|\epsilon}\notag\\
			=&&\sigma_{3}\frac{E_{3}+a^2 E_{3}-a J_{3}}{\left|a s_{3}-1\right|\epsilon}+\sigma_{4}\frac{E_{4}+a^2 E_{4}-a J_{4}}{\left|a s_{4}-1\right|\epsilon}+O(\epsilon).
	\end{eqnarray}
	As we have discussed in the above section, we could classify different orbits by the value of $b$. If $b=b_{cr}$, the orbit is critical which means the turning point of radial velocity is just on the horizon. We also have near critical orbit whose $b=b_{cr}+O(\epsilon)$ and non-critical orbit whose $b=b_{cr}+O(\epsilon^{0})$. In the following analysis, without loss of generality, along the same line of \cite{Maeda:2018hfi,r6}, the
calculation is carried out in case that particle 1 is critical ($J_1=b_{cr}E_1$), while particle 3 is near-critical($J_3=b_{cr}E_3 + O(\epsilon)$) and particle 2 is non-critical($J_2<b_{cr}E_2$) \cite{Maeda:2018hfi,r6}. Note that in our case $b_{cr}=\frac{a^2+1}{a}$, then we have
	\begin{eqnarray}
		&&E_{1}+a^2 E_{1}-a J_{1}=0,\\
		&&(\sigma_{4}\frac{1}{1-a s_{4}}-\sigma_{2}\frac{1}{1-a s_{2}})(a J_{2}-a^2 E_{2}- E_{2})=O(\epsilon).
	\end{eqnarray}

	Due to $J_{2}<b_{cr} E_{2}$, the particle 2 must be ingoing before the collision, which means $\sigma_{2}=-1$. We assume that the particle 4 is ingoing as well. Thus we should set $s_{1}=s_{3}$. To facility the calculations, we introduce parameters $\zeta<0, \alpha_{3}>0$ and $\beta_{3}$ as
	\begin{eqnarray}
		J_2&=&b_{cr}E_{2}(1+\zeta),\\
		J_3&=&b_{cr} E_3\left(1+\alpha_3\epsilon+\beta_3\epsilon^2+O(\epsilon^3)\right),
	\end{eqnarray}
	where $\alpha_3,\beta_3$ are the parameters of $O(\epsilon^{0})$. Next we shall calculate the expression of the energy of particle 3 and 2 through the conservative relations. Firstly, we expand the radial velocity of the four particles:
		\begin{eqnarray}
		u^{(1)}_{1}&=&\frac{\sigma_{1} f(E_{1},s_{1},a,0)}{a \left(1-a^2 s_{1}^2\right)}-\frac{E_{1}^2 \sigma_{1} \epsilon  h(s_{1},a)}{a \left(1-a^2 s_{1}^2\right)^2 f(E_{1},s_{1},a,0)}+O(\epsilon^2),\\
		u^{(1)}_{2}&=&-\frac{E_{2} \left(2 a^2 s_{2}+2 a-s_{2}\right) \left(a^2 \zeta -a s_{2}+\zeta +1\right)}{a (a s_{2}-1)^2 (a s_{2}+1)}+\frac{\left(a^2+1\right) E_{2} \zeta }{\epsilon  (1-a s_{2})}+\epsilon  \Pi +O(\epsilon^2),\\
		u^{(1)}_{3}&=&\frac{\sigma_3 f(E_3,s_1,a,\alpha_3)}{a \left(1-a^2 s_1^2\right)}-\frac{E_3^2 \sigma_3 \epsilon  (\beta_3 g_{1}(s_1,a,\alpha_3)-g_{2}(s_1,a,\alpha_3)+h(s_1,a))}{a \left(a^2 s_1^2-1\right)^2 f(E_3,s_1,a,\alpha_3)}+O(\epsilon^2),\nn\\
		u^{(1)}_{4}&=&\frac{\left(2 a^3 s_2^2-a \left(s_2^2+2\right)+s_2\right) (E_1+E_2-E_3)-\left(a^2+1\right) E_2 \zeta  (2 a (a s_2+1)-s_2)}{a (a s_2-1)^2 (a s_2+1)}\notag\\
		&&+\frac{\left(a^2+1\right) E_2 \zeta }{\epsilon  (1-a s_2)}+\frac{\left(a^3+a\right) \alpha_3 E_3 \left(a^2 s_2^2-1\right)}{a (a s_2-1)^2 (a s_2+1)}+\epsilon  \Omega+O(\epsilon^2),
	\end{eqnarray}
	where $\Omega$ and $\Pi$ are functions relevant to $E_{3},E_{2},s_{1},s_{2},b$ and $a$, which can be found in the appendix, and
\begin{eqnarray}
f(E,\text{s},a,\alpha_{3})&=&\sqrt{E^2 \left( \left(a^3+a\right) \alpha_{3} (a s+1)-a (2 a s+s+2)+s+1 \right) }  \notag\\
			&&\times \sqrt{ \left(a \left(\left(a^2+1\right) \alpha_{3} (a s+1)-2 a s+s-2\right)+s-1\right)-a^2  \left(a^2 s^2-1\right)^2},\\
h(\text{s},a)&=&\left(6 a^2 \left(a^2-1\right)^2-1\right) s^4+a \left(12 a^2-5\right) s+2 a^2+a \left(20 a^4-11 a^2+2\right) s^3\notag\\
			&&+\left(24 a^4-19 a^2+4\right) s^2-1,\\
g_{1}(\text{s},a,\alpha_{3})&=&a \left(a^2+1\right) (a s-1) (a s+1)^2 \left(a \left(a^2 \alpha_{3}+\alpha_{3}-2\right) (a s+1)+s\right),\\
g_{2}(\text{s},a,\alpha_{3})&=&\left(a^2+1\right) \alpha_{3} (a s_{1}+1)^2 \left(3 a^4 s_{1}^2+8 a^3 s_{1}+a^2 \left(5-3 s_{1}^2\right)-4 a s_{1}+s_{1}^2-1\right)\notag\\
			&&-a \left(a^2+1\right)^2 \alpha_{3}^2 (a s_{1}+1)^2 (2 a (a s_{1}+1)-s_{1}).
\end{eqnarray}
%which, in our following discussion, would be used to examine whether we can take certain value for these variables as needed.\\
The conservation equation demands
\begin{equation}
u^{(1)}_1+u^{(1)}_2=u^{(1)}_3+u^{(1)}_4.\label{radialvconservation}
\end{equation}
For $\zeta$ is negative, we expand Eq. \eqref{radialvconservation} order by order, and found that the identity relation is always satisfied for the order of $\epsilon^{-1}$. Then we consider the order of $\epsilon^{0}$ and get
	\begin{equation}
		\frac{\sigma_{1} f(E_{1},s_{1},a,0)}{a \left(1-a^2 s_{1}^2\right)}-\frac{E_{3} g_{1}(s_{2},a,\alpha_{3})}{a \left(1-a^2 s_{2}^2\right)}-\frac{\sigma_{3} f(s_{1},E_{3},\alpha_{3})}{a \left(1-a^2 s_{1}^2\right)}+\frac{E_{1} \left(2 a^2 s_{2}+2 a-s_{2}\right)}{a \left(1-a^2 s_{2}^2\right)}=0.
	\end{equation}
	It can be seen as a quadratic equation of $E_3$ as
	\begin{equation}
		\label{E3}
		A_{m} E_{3}^2-2 B_{m} E_{3}+C_{m}=0,
	\end{equation}
	where the detailed expression $A_{m}, B_{m}$ and $C_{m}$ can be found in the appendix.
	
	The conservation equation of the order of $\epsilon^{0}$ can also be written as
	\begin{eqnarray}
		&&E_{3}+\frac{\sigma_{3} f(E_{3},s_{1},a,\alpha_{3})a (1-a^2 s_{2}^2)}{a (1-a^2 s_{1}^2) g_{1}(s_{2},a,\alpha_{3})}\notag\\
		= &&\frac{1}{g_{1}(s_{2},a,\alpha_{3})}\left[E_{1} \left(2 a^2 s_{2}+2 a-s_{2}\right)+\frac{\sigma_{1} f(E_{1},s_{1},a,0)a (1-a^2 s_{2}^2)}{a \left(1-a^2 s_{1}^2\right)}\right].
	\end{eqnarray}

	The right side of the equation can be considered as a constant irrelevant to $E_{3}$. Hence the range of $E_{3}$ is affected by $\sigma_{3}$. When $\sigma_{3}$ is negative, $E_{3}$ must be larger than the value of the right side. The larger
solution of $E_3$ gives larger efficiency because the efficiency depends on the value of
$E_3$ \cite{Maeda:2018hfi}. Therefore, it is sufficient to consider the case
of the negative value of $\sigma_{3}$.
	
	We can solve the equation \eqref{E3} and only retain the larger one
	\begin{equation}
		E_{3}=\frac{\sqrt{B_{m}^2-A_{m} C_{m}}+B_{m}}{A_{m}}.
\label{nihao}
	\end{equation}
	While $E_{3}$ is fixed, we can also get the value of $E_{2}$ for the given $E_{1},a,s_{1},s_{2},\alpha_{3}$ and $\beta_{3}$ through the conservation equation of radial momentum in the order of $\epsilon^{1}$. The efficiency of energy contraction is defined by
	\begin{equation}
		\eta =\frac{E_{3}}{E_{1}+E_{2}}.
	\end{equation}
	Thus we need only care about whether $E_{2}$ can be taken the smallest value of 1. That is what we are going to analyze in the next section.
	%%%%%%%%%%%%%%%%%%%%%%%%%%%%%%%%%%%%%%%%%%%%%%%%%%%%%%%%%%%%
	%%%%%%%%%%%%%%%%%%%%%%%%%%%%%%%%%%%%%%%%%%%%%%%%%%%%%%%%%%%%
	\section{THE MAXIMAL EFFICIENCY}
	The solution to Eq.\eqref{E3} already gives the value of $E_{3}$, which depends on the parameter $E_{1},a,s_{1},s_{2}$ and $\alpha_{3}$. We normalize $E_{1}=1$ and $\sigma_{1}=1$, instead of $\sigma_{1}=-1$ which can not get higher efficiency \cite{Leiderschneider:2015kwa}. Hence, if we know the value of $d$ (or $a)$, we can find a maximal value of $E_{3}$ in terms of $s_{1},s_{2}$ and $\alpha_{3}$ under the time-like condition.
	
\begin{figure}
	\centering
\subfigure{
\begin{minipage}{3.5cm}
\centering
\includegraphics[width=3.5cm]{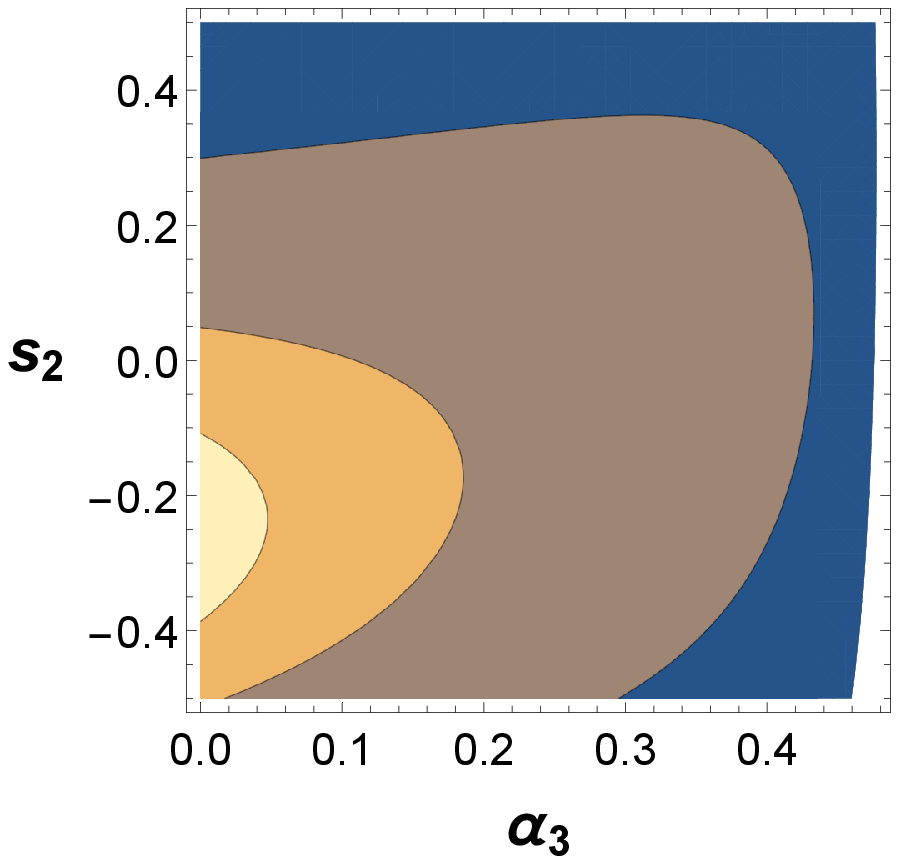}
\begin{center}
$($a$)$ $d=0.08$
\end{center}
\end{minipage}
}
\subfigure{
\begin{minipage}{0.3cm}
\centering
\includegraphics[height=3.5cm,width=0.3cm]{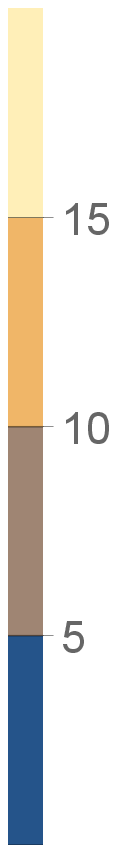}
\begin{center}

\end{center}
\end{minipage}
}
\subfigure{
\begin{minipage}{3.5cm}
\includegraphics[width=3.5cm]{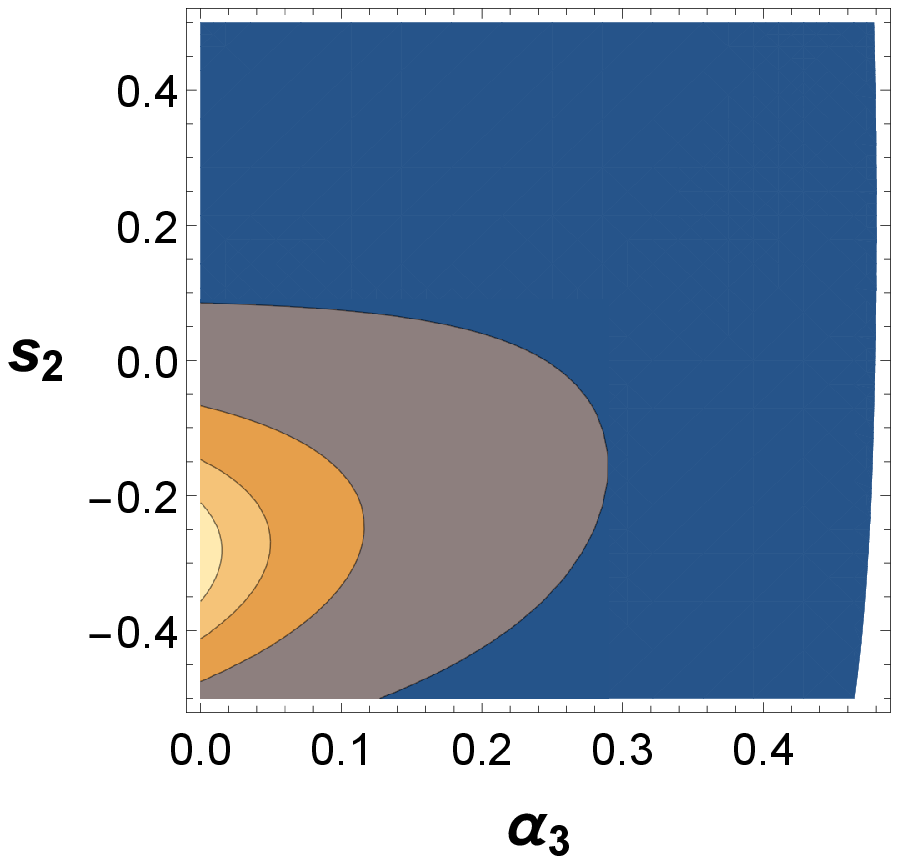}
\begin{center}
$($b$)$ $d=-0.12$
\end{center}
\end{minipage}
}
\subfigure{
\begin{minipage}{0.3cm}
\centering
\includegraphics[height=3.5cm,width=0.3cm]{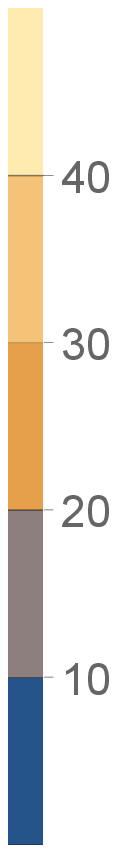}
\begin{center}

\end{center}
\end{minipage}
}
\subfigure{
\begin{minipage}{3.5cm}
\includegraphics[width=3.5cm]{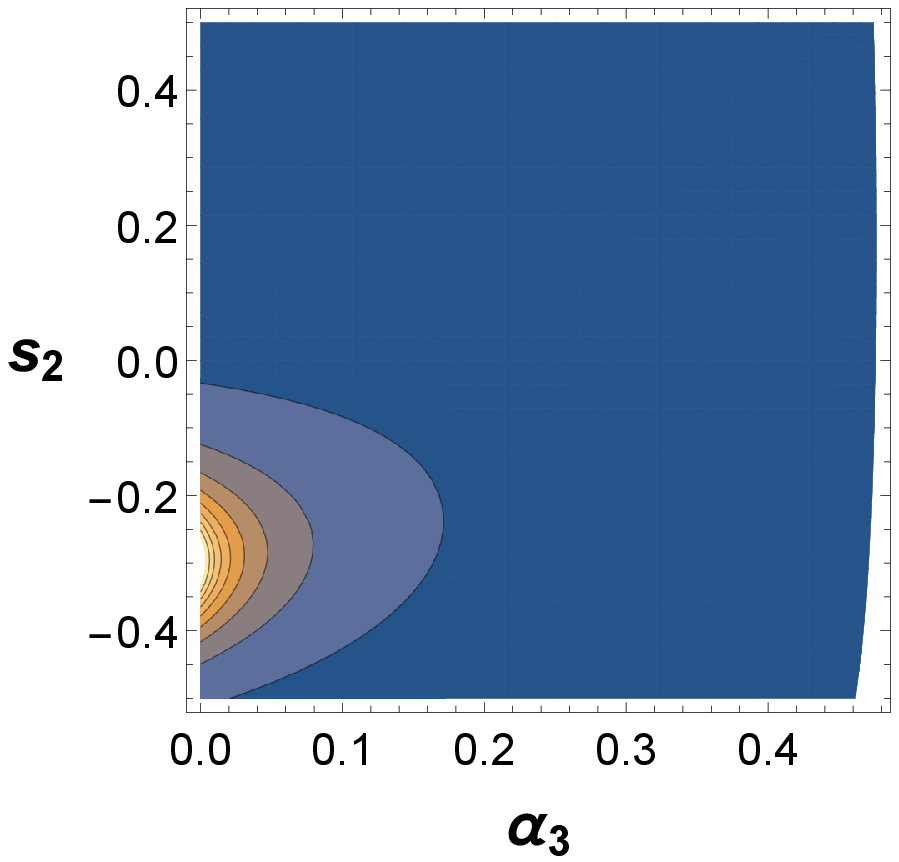}
\begin{center}
$($c$)$ $d=-0.24$
\end{center}
\end{minipage}
}
\subfigure{
\begin{minipage}{0.3cm}
\centering
\includegraphics[height=3.5cm,width=0.3cm]{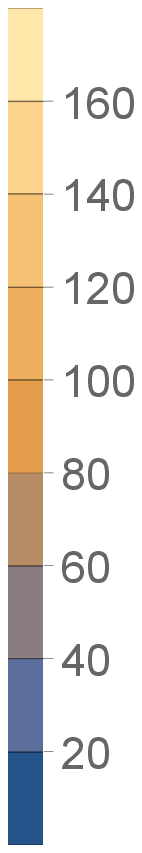}
\begin{center}

\end{center}
\end{minipage}
}
\caption{ From the contour map of $E_{3}$ we find for distinct $d$.
}
\label{Fig4}
\end{figure}

\begin{figure}[htbp]
	\centering
\subfigure{
\begin{minipage}{3.5cm}
\centering
\includegraphics[width=3.5cm]{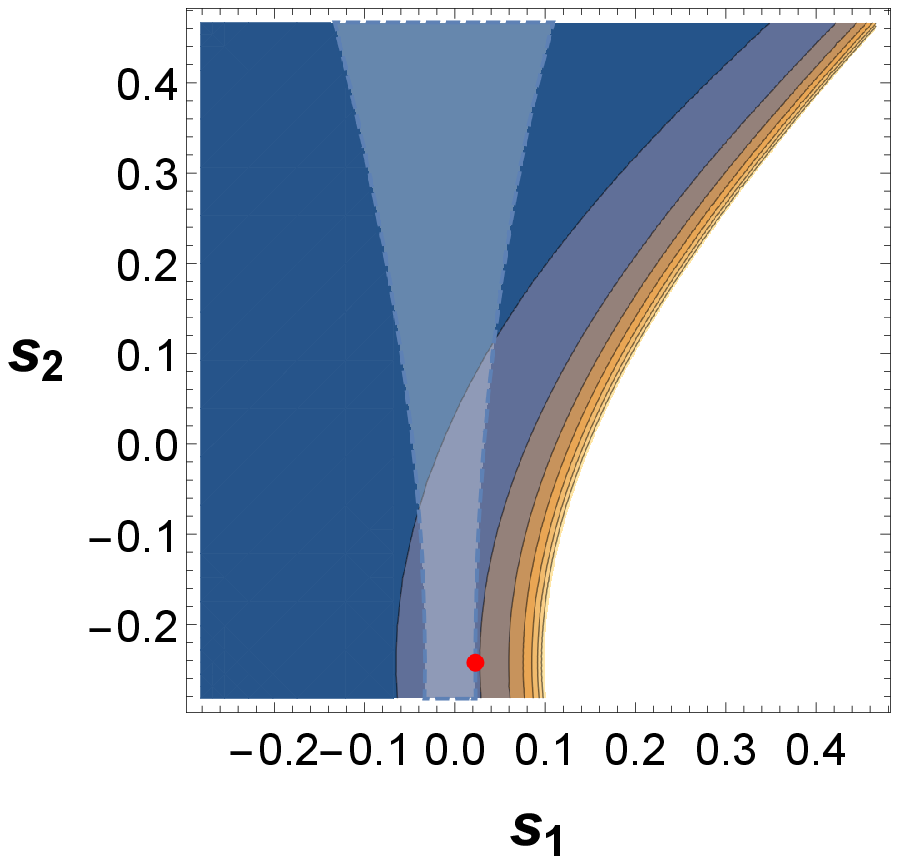}
\begin{center}
$($a$)$ $d=0.12$
\end{center}
\end{minipage}
}
\subfigure{
\begin{minipage}{0.3cm}
\centering
\includegraphics[height=3.5cm,width=0.3cm]{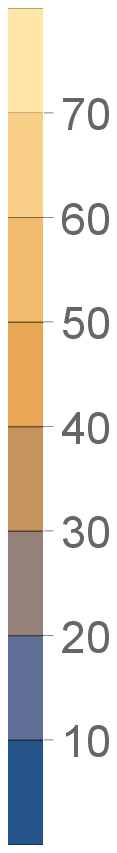}
\begin{center}

\end{center}
\end{minipage}
}
\subfigure{
\begin{minipage}{3.5cm}
\includegraphics[width=3.5cm]{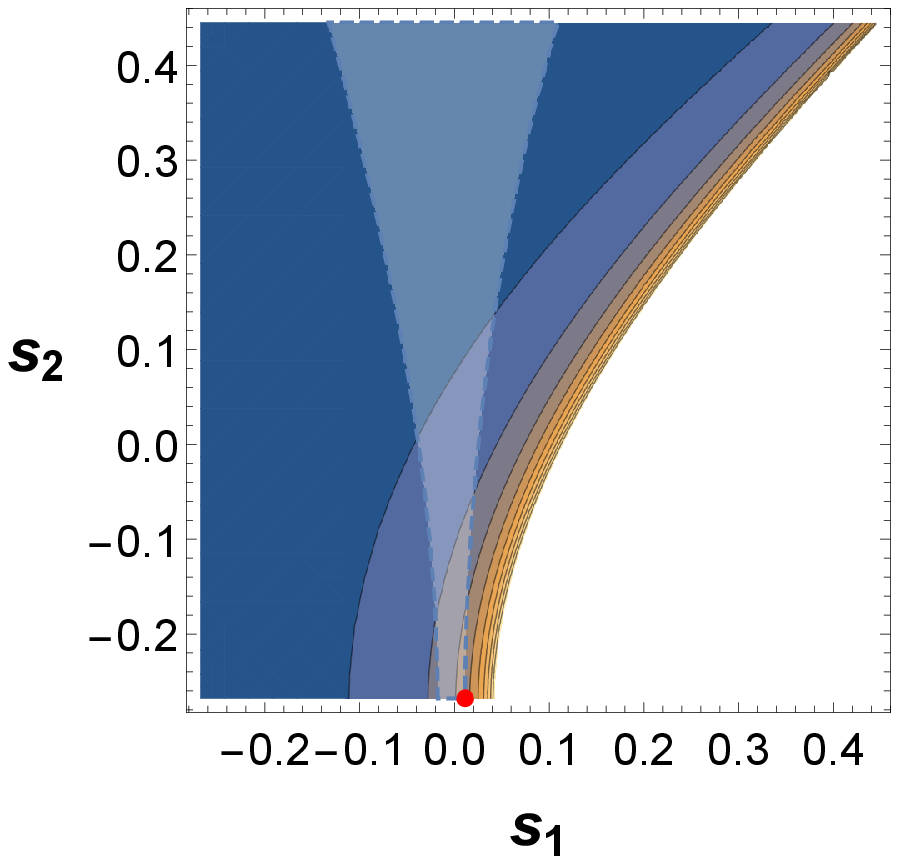}
\begin{center}
$($b$)$ $d=-0.04$
\end{center}
\end{minipage}
}
\subfigure{
\begin{minipage}{0.3cm}
\centering
\includegraphics[height=3.5cm,width=0.3cm]{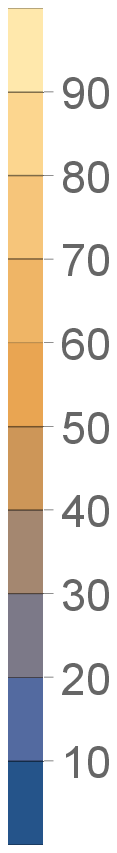}
\begin{center}

\end{center}
\end{minipage}
}
\subfigure{
\begin{minipage}{3.5cm}
\includegraphics[width=3.5cm]{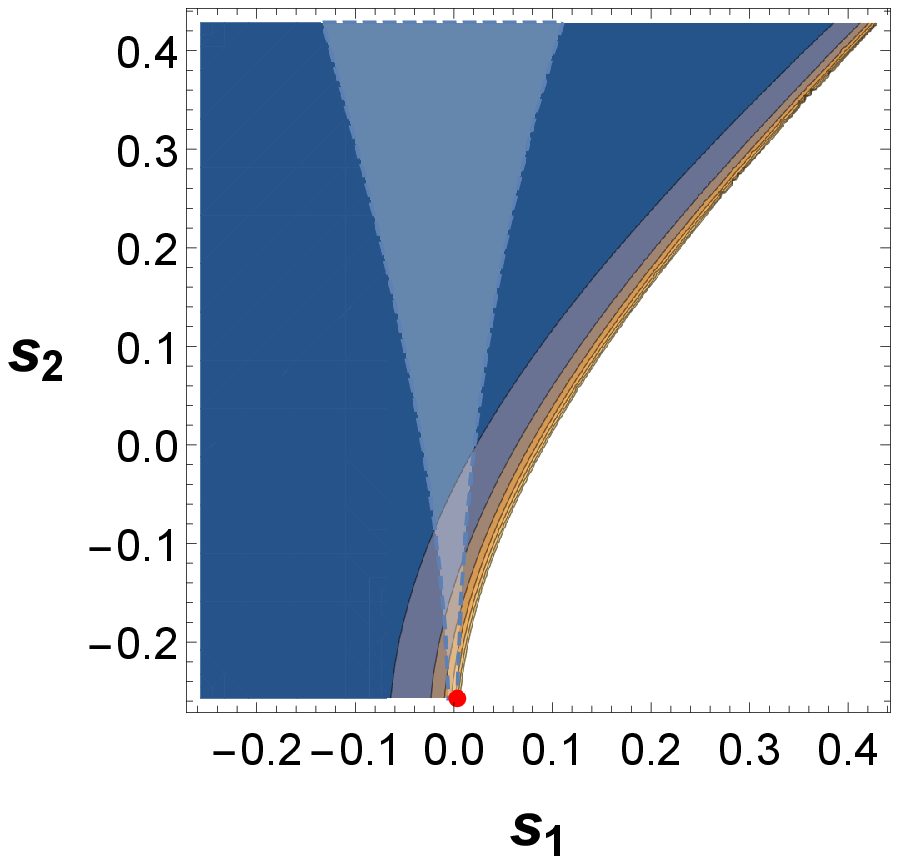}
\begin{center}
$($c$)$ $d=-0.20$
\end{center}
\end{minipage}
}
\subfigure{
\begin{minipage}{0.3cm}
\centering
\includegraphics[height=3.5cm,width=0.3cm]{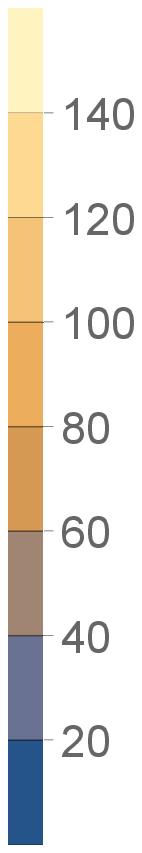}
\begin{center}

\end{center}
\end{minipage}
}
\caption{The point with maximal $E_{3}$ as the function of $s_{1}$ and $s_{2}$ under the time-like condition is marked in red in the contour map.}
\label{Fig5}
\end{figure}

	We plot the contour map of $E_{3}$ in Fig. \ref{Fig4} for distinct $d$ and we find that $E_{3}$ can always achieves the maximal value in the condition $\alpha_{3}=0$.
We also plot the contour map of $E_{3}$ as the function of $s_{1}$ and $s_{2}$ under the time-like condition in Fig. \ref{Fig5} for different values of $d$.

\begin{figure}
\centering
\includegraphics[width=8cm]{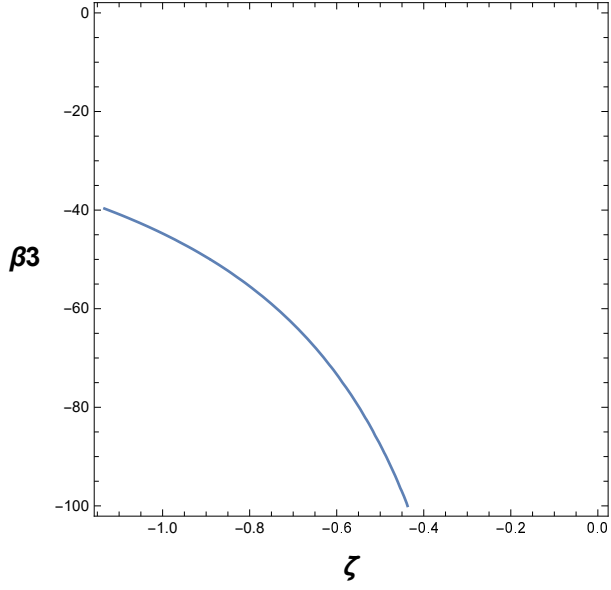}
\caption{The relationship between  $\beta_{3}$ and $\zeta$ for $d=-0.24$ when $E_{3}$ is fixed $($the maximal value$)$ under timelike condition. It reflects that $E_{2}=1$ can be satisfied in this case.}
\label{Fig6}
\end{figure}

	For each $d$, we could find a maximal value of $E_{3}$. Note that $E_2\geq 1$ when the particle 2 falling from infinity, if $E_2=1$ is possible, then the maximal value of $E_3$ gives the maximal efficiency. Note that $E_3$ is decoupled with
	parameters $\beta_3$ and $\xi$. So our target is simplified to sure $E_2=1$ is possible for some admissible values
	of $\beta_3$ and $\xi$. In what follows we should focus on whether $E_{2}$ can be 1 when $E_{3}$ is taken the maximal value. Note that Eq.\eqref{xi} gives the constraints on $\zeta$, then we set $E_{2}=1$ and observe whether there exists some value of $\beta_{3}$ in the given range of $\zeta$. We show the results in Fig. \ref{Fig6} when $d=-0.24$. The result for other values of $d$ is similar, and we found that $E_{2}=1$ can always be achieved.
	
\begin{figure}
\centering
\includegraphics[width=8cm]{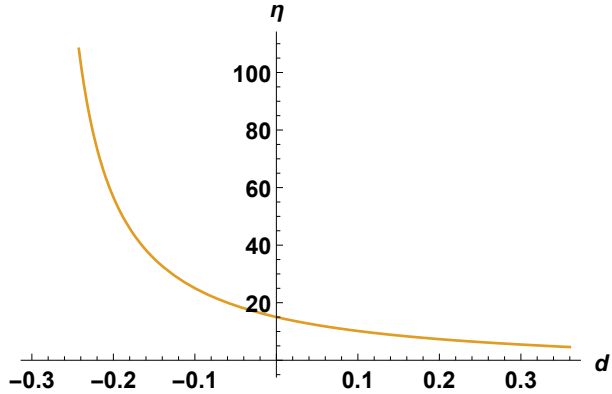}
\caption{The maximal efficiency of $\eta$ as a function of $d$.}
\label{Fig7}
\end{figure}

Therefore we list the maximal efficiency for different $d$ in Table \ref{Table1}. In particular, when $d=0$, the spacetime metric reduces to the Kerr black hole, the maximal efficiency also goes back to 15.01 which is completely the same as in\cite{Maeda:2018hfi}. Moreover, for the negative value of $d$, the efficiency $\eta$ is larger than the Kerr case. The maximal efficiency $\eta$ as a function of $d$ is plotted in Fig. \ref{Fig7}. The result shows that $\eta$ increases as the parameter $d$ decreases.
	
\begin{table}
  \centering
  \begin{tabular}{|c|c|c|c|c|c|c|}
\hline
$d$ & $a$ & $s_{1}$ & $s_{2}$ & $E_{3}$ &$E_{2}$& $\eta$ \\
\hline
-0.24&	1.11355&	0.0019225&	-0.254649&	206.255&	1&	103.127\\
\hline
-0.20&	1.09545&	0.00351892&	0.256994&	113.359&	1&	56.6796\\
\hline
-0.12&	1.0583&	0.00710832&	-0.262068&	56.8549&	1&	28.4274\\
\hline
0.00&	1&	0.0137838&	-0.268102&	30.0203&	1&	15.0101\\
\hline
0.12&	0.938083&	0.0224733&	-0.242029&	18.976&	1&	9.48799\\
\hline
0.24&	0.87178&	0.0341456&	-0.202261&	13.0032&	1&	6.50159\\
\hline
    \end{tabular}
\caption{The maximal efficiency and its corresponding value of $s_{1},s_{2}$ and $E_{2}$ for different $d$}
\label{Table1}
\end{table}

	%%%%%%%%%%%%%%%%%%%%%%%%%%%%%%%%%%%%%%%%%%%%%%%%%%%%%%%%%%%%
	%%%%%%%%%%%%%%%%%%%%%%%%%%%%%%%%%%%%%%%%%%%%%%%%%%%%%%%%%%%%
	\section{CONCLUDING REMARK}

In this paper, we study the collision of two massive spinning particles around an
extreme braneworld black hole and calculate the corresponding maximal efficiency of the energy extraction
from the black hole which is usually referred to as the collisional Penrose process. Compared with Kerr spacetime, the braneworld background possesses an extra tilde charge parameter $d$. It reduces to the Kerr black hole when
the parameter $d=0$. We consider the particles freely falling from infinity and collide near the horizon of the braneworld
black hole. After a collision, one particle escapes to infinity with
larger energy than the incident particles.

By employing the MPD equations which describe the motion of the spinning particles in the braneworld spacetime, the momentum and the velocity of the particles have been obtained explicitly. Moreover, by imposing
timelike constraint which comes from the causality requirement, we found a restriction on the angular momentum of the particles. Then, with the help of the conservation of energy and momentum during the collision, we
calculate the corresponding maximum energy extraction efficiency $\eta_{max}$ of the collisional Penrose process,
and the detailed calculation shows the $\eta_{max}$ does depend on the tilde charge parameter $d$ of the braneworld spacetime.
When $d=0$, the corresponding maximum energy extraction efficiency $\eta_{max}=15.01$ goes back to the Kerr case \cite{Maeda:2018hfi}. This of course can be viewed as a consistency check. Moreover, for the negative value of $d$, the efficiency $\eta$ can be larger than the Kerr case. The result shows that $\eta_{max}$ increases as the parameter $d$ decreases.

	%%%%%%%%%%%%%%%%%%%%%%%%%%%%%%%%%%%%%%%%%%%%%%%%%%%%%%%%%%%
	%%%%%%%%%%%%%%%%%%%%%%%%%%%%%%%%%%%%%%%%%%%%%%%%%%%%%%%%%%%
	\section{ACKNOWLEDGEMENT}

		This work is supported by NSFC with No.11775082. Y. Du would like to thank Tianyi Li for helpful discussions.
	
	%%%%%%%%%%%%%%%%%%%%%%%%%%%%%%%%%%%%%%%%%%%%%%%%%%%%%%%%%%%%%%%%%%%%%%%%%%%%%%%%%%%%
	\appendix
	\par
	\newpage
	\section{appendix}
	\begin{eqnarray}
		A_{m}&=&\frac{g_{1}(s_{2},a,\alpha_{3})^2}{a^2 \left(a^2 s_{2}^2-1\right)^2}+\frac{-4 a^4 s_{1}^2-8 a^3 s_{1}+5 a^2 s_{1}^2-\left(a^2+1\right)^2 a^2 \alpha_{3}^2 (a s_{1}+1)^2}{a^2 \left(1-a^2 s_{1}^2\right)^2}\notag\\
		&&+\frac{2 \left(a^2+1\right) a \alpha_{3} (a s_{1}+1) (2 a (a s_{1}+1)-s_{1})-4 a^2+2 a s_{1}-s_{1}^2+1}{a^2 \left(1-a^2 s_{1}^2\right)^2},\\
		B_{m}&=&\frac{g_{1}(s_{2},a,\alpha_{3})E_{1} \left(2 a^2 s_{2}+2 a-s_{2}\right)}{a^2 \left(1-a^2 s_{2}^2\right)^2} +\frac{\sigma_{1} \sqrt{\mathcal{P}}}{a^2 \left(1-a^2 s_{1}^2\right) \left(1-a^2 s_{2}^2\right)},\\
		C_{m}&=&\left(\frac{E_{1} \left(2 a^2 s_{2}+2 a-s_{2}\right)}{a \left(1-a^2 s_{2}^2\right)}\right)^2+\left(\frac{\sigma_{1} \sqrt{\mathcal{P}}}{a \left(1-a^2 s_{1}^2\right)}\right)^2\notag\\
		&&+\frac{\sqrt{\mathcal{P}}  E_{1} \left(2 a^2 s_{2}+2 a-s_{2}\right)\sigma_{1} }{a^2 \left(1-a^2 s_{1}^2\right) \left(1-a^2 s_{2}^2\right)} +1.
	\end{eqnarray}
	where
	\begin{eqnarray}
		\mathcal{P}&=&a^6 \left(-s_{1}^4\right)+4 a^4 E_{1}^2 s_{1}^2+2 a^4 s_{1}^2+8 a^3 E_{1}^2 s_{1}-5 a^2 E_{1}^2 s_{1}^2+4 a^2 E_{1}^2-a^2-2 a E_{1}^2 s_{1}\notag\\
		&&+E_{1}^2 s_{1}^2-E_{1}^2
	\end{eqnarray}

\begin{eqnarray}
	\Omega&&=\notag\\
	&&\frac{\sigma_4}{\mathcal{H}_1 }\Big(
	a^6 s_2^2 \big(-12 s_2 (E_1+E_2-E_3) (J_1+J_2-J_3)+6 s_2^2 (E_1+E_2-E_3)^2\notag\\
	&&+14 (E_1+E_2-E_3)^2- s_2 ^4\big)+2 a^5  s_2  \big(-3  s_2 ^3 ( E_1+E_2-E_3  ) ( J_1+J_2-J_3  )\notag\\
	&&+ s_2 ^2 \left(7 ( E_1+E_2-E_3  )^2+3 ( J_1+J_2-J_3  )^2\right)-14  s_2  ( E_1+E_2-E_3  ) ( J_1+J_2-J_3  )\notag\\
	&&+5 ( E_1+E_2-E_3  )^2+ s_2 ^4\big)+a^4 \big(-8  s_2 ^3 ( E_1+E_2-E_3  ) ( J_1+J_2-J_3  )\notag\\
	&&+ s_2 ^2 \left(19 ( E_1+E_2-E_3  )^2+14 ( J_1+J_2-J_3  )^2\right)-20  s_2  ( E_1+E_2-E_3  ) ( J_1+J_2-J_3  )\notag\\
	&&+2 ( E_1+E_2-E_3  )^2+ s_2 ^4\big)+2 a^3 \big(3  s_2 ^4 ( E_1+E_2-E_3  ) ( J_1+J_2-J_3  )\notag\\
	&&+ s_2 ^3 \left(2 ( E_1+E_2-E_3  )^2-3 ( J_1+J_2-J_3  )^2-2\right)-7  s_2 ^2 ( E_1+E_2-E_3  ) ( J_1+J_2-J_3  )\notag\\
	&&+ s_2  \left(4 ( E_1+E_2-E_3  )^2+5 ( J_1+J_2-J_3  )^2\right)-2 ( E_1+E_2-E_3  ) ( J_1+J_2-J_3  )\big)\notag\\
	&&+a^2 \big(8  s_2 ^3 ( E_1+E_2-E_3  ) ( J_1+J_2-J_3  )+ s_2 ^2 \left(4 ( E_1+E_2-E_3  )^2-5 ( J_1+J_2-J_3  )^2+1\right)\notag\\
	&&-4  s_2  ( E_1+E_2-E_3  ) ( J_1+J_2-J_3  )-5  s_2 ^4 ( E_1+E_2-E_3  )^2+( E_1+E_2-E_3  )^2\notag\\
	&&+2 ( J_1+J_2-J_3  )^2\big)+6 a^7  s_2 ^3 ( E_1+E_2-E_3  )^2+2 a  s_2  \big( s_2 ^3 (-( E_1+E_2-E_3  )) ( J_1+J_2-J_3  )\notag\\
	&&+3  s_2  ( E_1+E_2-E_3  ) ( J_1+J_2-J_3  )-3 ( E_1+E_2-E_3  )^2+ s_2 ^2 ( J_1+J_2-J_3  )^2\notag\\
	&&-2 ( J_1+J_2-J_3  )^2+1\big)- \left( s_2  \left(2  s_2 ^2-7\right) ( E_1+E_2-E_3  )+ J_1+J_2-J_3  \right)\notag\\
	&&\times (- s_2  ( E_1+E_2-E_3  )+ J_1+J_2-J_3  )-1
	\Big)
\end{eqnarray}

where
\begin{eqnarray}
	\mathcal{H}_1=2 \sqrt{\left(E_1 a^2+E_{2} a^2-E_{3} a^2-J_{1} a-J_{2} a+J_{3} a+E_1+E_{2}-E_{3}\right)^2} (a s_{2}-1) (a s_{2}+1)^2
\end{eqnarray}

	\begin{eqnarray}
		\Pi&=\notag\\
		&&\frac{\sigma_2}{\mathcal{H}_2 }
		\big(		
		\left(a^2+1\right)^2  E_2 ^2 \zeta ^2 \left(a \left(2 \left(3 a^4-3 a^2+1\right)  s_2 ^3+a \left(14 a^2-5\right)  s_2 ^2+2 \left(5 a^2-2\right)  s_2 +2 a\right)-1\right)\notag\\
		&&-2 \left(a^2+1\right)  E_2 ^2 \zeta  \left(a \left(\left(3 a^4-3 a^2+1\right)  s_2 ^3+a \left(7 a^2-3\right)  s_2 ^2+\left(5 a^2-1\right)  s_2 +a\right)-1\right) (a  s_2 -1)\notag\\
		&&-(a  s_2 -1)^4 \left(a^2 (a  s_2 +1)^2+ E_2 ^2\right)
		\big)
	\end{eqnarray}
where
	\begin{eqnarray}
		\mathcal{H}_2=2 a^2 (a  s_2 -1)^4 (a  s_2 +1)^2 \sqrt{\frac{\left(a^2+1\right)^2  E_2 ^2 \zeta ^2}{(a  s_2 -1)^2}};
	\end{eqnarray}
\newpage

	\bibliographystyle{unsrt}

\end{document}